%% file: main.tex
\documentclass{article}

     \usepackage[final]{neurips_2019}

\usepackage[utf8]{inputenc} 
\usepackage[T1]{fontenc}    
\usepackage{hyperref}       
\usepackage{url}            
\usepackage{booktabs}       
\usepackage{amsfonts}       
\usepackage{nicefrac}       
\usepackage{microtype}      

\usepackage{amsmath}
\usepackage{amssymb}
\usepackage{graphicx}
\usepackage{sidecap}
\sidecaptionvpos{figure}{c}
\usepackage{wrapfig}
\graphicspath{ {./figures/} }
\usepackage{caption}
\usepackage{subcaption}
\usepackage{siunitx}
\usepackage[table]{xcolor}
\definecolor{LightCyan}{rgb}{0.88,1,1}

\title{Learning from Data-Rich Problems: A Case Study on Genetic Variant Calling}
\author{%
    Ren Yi\thanks{This work was done while RY was with Google.}\\
    Department of Computer Science\\
    New York University\\
    New York, NY 10011 \\
    \texttt{ren.yi@nyu.edu}
    \And
    Pi-Chuan Chang, Gunjan Baid \& Andrew Carroll \\
    Google\\
    3400 Hillview Avenue\\
    Palo Alto, CA 94304\\
    \texttt{ \{pichuan,gunjanbaid,awcarroll\}@google.com}
    }
    
\begin{document}

\maketitle

\begin{abstract}
    Next Generation Sequencing can sample the whole genome (WGS) or the 1-2\% of the genome that codes for proteins called the whole exome (WES). Machine learning approaches to variant calling achieve high accuracy in WGS data, but the reduced number of training examples causes training with WES data alone to achieve lower accuracy. We propose and compare three different data augmentation strategies for improving performance on WES data: 1) joint training with WES and WGS data, 2) warmstarting the WES model from a WGS model, and 3) joint training with the sequencing type specified. All three approaches show improved accuracy over a model trained using just WES data, suggesting the ability of models to generalize insights from the greater WGS data while retaining performance on the specialized WES problem. These data augmentation approaches may apply to other problem areas in genomics, where several specialized models would each see only a subset of the genome.
\end{abstract}

\section{Introduction}
\input{introduction.tex}

\vspace{-10pt}
\section{Related Work}\label{sec:related_work}
\input{related_work.tex}

\section{Approaches}
\input{method.tex}

\section{Experiments}
\input{experiment.tex}

\section{Results}
\input{result.tex}

\vspace{-8pt}
\section{Conclusion}
\input{conclusion.tex}

\subsubsection*{Acknowledgments}
We thank Babak Alipanahi, Cory McLean, Sidharth Goel, Alexey Kolesnikov, Maria Nattestad, Thomas Colthurst, Ming-Wei Chang, and other colleagues at Google for providing helpful suggestions on the manuscript and the project in general.

\bibliography{reference.bib}
\bibliographystyle{unsrtnat}

\end{document}

%% file: introduction.tex
Next Generation Sequencing (NGS) measures a genome by repeated, semi-random sampling of short (76-300bp) fragments that have a 1\% base error rate. NGS can sample the whole genome (WGS) or can attempt to target coverage to the whole exome (WES), the 1-2\% of genome which codes for proteins and their bordering regions. WES is a cost-effective method for identifying interpretable, causal variants in Mendelian disorders~\citep{bamshad2011exome}.

After sequencing, variant calling analyzes these fragments relative to a reference genome to identify the genomic positions that distinguish an individual sample~\citep{GATK,freebayes,Clairvoyante,strelka2}. Machine learning approaches to variant calling~\citep{Clairvoyante,deepVariant,campagnelab} have demonstrated best-in-class accuracy, benefitting from training sets created by extensively sequencing the well-characterized Genome in a Bottle (GIAB) samples~\citep{giab_data,benchmark}.
 
WES must contend with greater sources of error (e.g. variation in capture efficiency and greater GC bias~\citep{wgsBetterWes}), and WES samples generate less training data since WES covers less of the genome. Although there are a great deal of publicly available WES data, very few of them are generated on the GIAB truth sets needed for training and evaluating variant calling models. In this work, we investigate approaches that allow machine learning to benefit from the substantially larger body of WGS training data while retaining specialized learning from WES training data.
 
\begin{figure}[!ht]
\centering
\includegraphics[width=\textwidth]{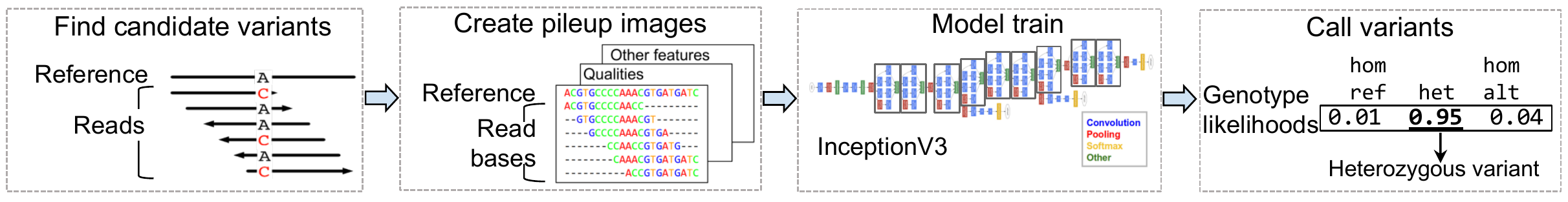}
\caption{The DeepVariant workflow.}
\label{fig:dv_architecture}
\end{figure}

We use DeepVariant~\citep{deepVariant} as the foundation for this investigation. DeepVariant performs variant calling in four steps: 1) scanning through NGS read alignments to find candidate variants, 2) local reassembly of reads to reference and candidate variant haplotypes, 3) creation of a six-channel pileup image that represents the bases, base quality, mapping quality, strand, and support for reference or variant haplotype over a 221bp window, and 4) using an InceptionV3~\citep{inceptionv3} deep neural net to predict the genotype at the candidate position (Figure~\ref{fig:dv_architecture}).
 
\begin{SCfigure}[][!ht]
     \centering
     \caption{DeepVariant performance using only WES or WGS data. F1 scores are calculated based on single nucleotide polymorphisms (SNPs) and indels (insertions and deletions) prediction accuracy. The experimental dataset used for the comparison is described in Section~\ref{sec:data}.}
     \includegraphics[width=0.5\textwidth]{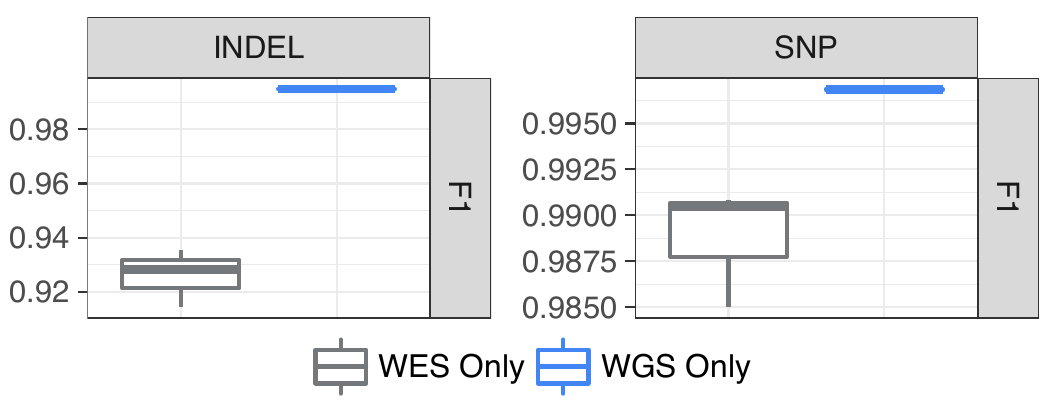}
     \label{fig:single_datatype}
\end{SCfigure}

\vspace{-10pt}
\begin{wraptable}{r}{6.5cm}
\centering
\begin{tabular}[t]{lrr}
\toprule
& \multicolumn{1}{c}{WGS} & \multicolumn{1}{c}{WES}\\
\midrule
Train & $320,662,815$ & $17,402,861$\\
Tune & $2,435,712$ & $631,261$\\
\bottomrule
\end{tabular}
\caption{The number of examples in DeepVariant production datasets.}
\label{tab:example_count}
\end{wraptable}
Currently, separate DeepVariant models are trained for WGS and WES data, termed WGS and WES models, respectively. However, due to the inherent low region coverage, the exome contains far fewer variants ($2\times10^5$) than the genome ($4\times10^6$), resulting in far fewer training examples from WES data compared to WGS (Table~\ref{tab:example_count}). Variant calling performance (measured by F1 score) achieved using only the WES data is also considerably lower and less stable than that of the WGS (Figure~\ref{fig:single_datatype}).

%% file: related_work.tex
Deep neural networks require large amounts of data to achieve high accuracy in computer vision~\citep{imagenet,cifar10}, natural language processing~\citep{squad,MultiNLI}, and genomics~\citep{encode,gencode,remap} tasks. Data augmentation techniques borrow from data-rich problems or generate adversarial examples. Image augmentation generates new examples by adding random noise and transformations to existing images~\citep{augmentimage,autoaugment}. This process is extended by generative adversarial networks~\citep{gans,stylegan,dagan}, which are especially useful for highly-skewed data and uncommon cases~\citep{gan4med}.

Few methods have been proposed to generate adversarial examples for variant calling. The incomplete understanding of sequencing error profile and genome content forces strategies to semi-simulate data~\citep{semicall}, but the faithfulness with which these approximate real-world data has not been comprehensively evaluated.

%% file: method.tex
\begin{wrapfigure}{r}{0.4\textwidth}
\vspace{-16pt}
     \includegraphics[width=\linewidth]{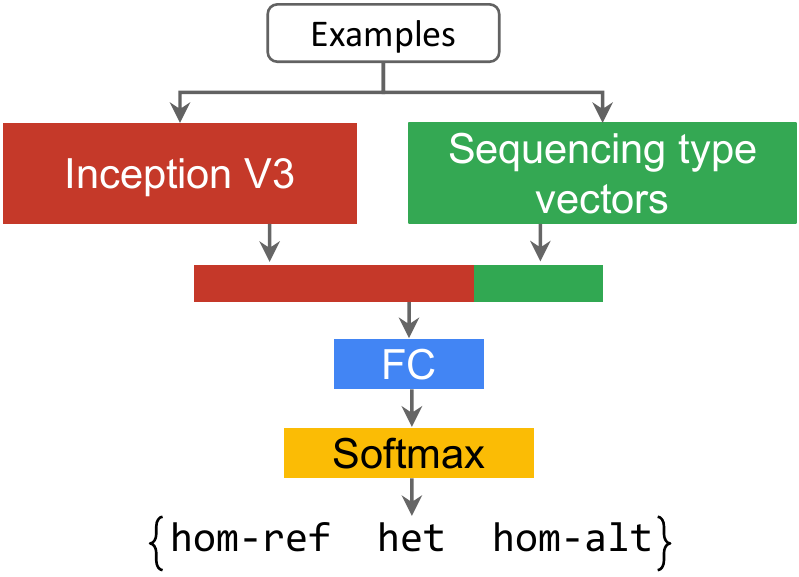}
     \caption{SeqType architecture.}
     \label{fig:model}
     \vspace{-13pt}
\end{wrapfigure}
We establish the baseline for training WES models--training from WES data alone (\textit{WES Only}).

We first investigate two naive strategies for adding WGS training examples: 1) training a model from a combination of WGS and WES data (\textit{WGS + WES}), and 2) warmstarting a WES model from a trained WGS model (\textit{warmstart WGS}).

Additionally, we introduce an additional low-dimensional vector to DeepVariant to capture sequencing types (\textit{SeqType}). In the \textit{SeqType} approach (Figure~\ref{fig:model}), we associate each example pileup image with its sequencing type, which is a randomly initialized vector for the two data types (WES or WGS). This vector is concatenated with the output of the InceptionV3 PreLogit layer to form the final feature vector, which is then provided to a feedforward network followed by a softmax layer to produce the final genotype probabilities.

%% file: experiment.tex
\subsection{Data}\label{sec:data}

We use a reduced set of DeepVariant’s production dataset to minimize data heterogeneity (Table~\ref{tab:reduced_example_count}). This experimental dataset contains three PCR-free WGS BAM files sequenced on Illumina HiSeq2500 and 18 WES BAM files sequenced on Illumina HiSeq4000.

\begin{wraptable}{r}{6cm}
\vspace{-10pt}
\begin{tabular}[t]{lrr}
\toprule
& \multicolumn{1}{c}{WGS} & \multicolumn{1}{c}{WES}\\
\midrule
Train & $37,106,930$ & $2,641,013$\\
Tune & $1,024,080$ & $94,149$\\
\bottomrule
\end{tabular}
\caption{The number of examples proposed by DeepVariant using the experimental dataset.}
\label{tab:reduced_example_count}
\vspace{-15pt}
\end{wraptable}
The GIAB truth sets~\citep{giab_data,benchmark} provide labels for training and evaluation. We use HG001 samples for training and hold out HG002 for evaluation. This is the same training and evaluation strategy used for DeepVariant. The training set for HG001 is the v3.3.2 truth set, while the evaluation set for HG002 uses the v4-beta truth set newly available for only this sample~\citep{giab_data,benchmark}.

\subsection{Experimental setup}

For each experiment, the checkpoint that achieves the highest F1 score on the tuning set within the first 2 million steps is selected as the best model checkpoint. The experiments are performed on TPUs~\citep{tpu}. We follow the DeepVariant WES case study\footnote{\url{https://github.com/google/deepvariant/blob/r0.8/docs/deepvariant-exome-case-study.md}} to evaluate the model performance using fully held-out HG002 WES sample available from GIAB~\citep{giab_data}. Variant predictions are bootstrapped 100 times. These bootstrap samples are used to perform statistical analyses, and $p$-values are calculated based on student's t-test.

%% file: result.tex
\begin{figure}[!ht]
\centering
\includegraphics[width=0.75\textwidth]{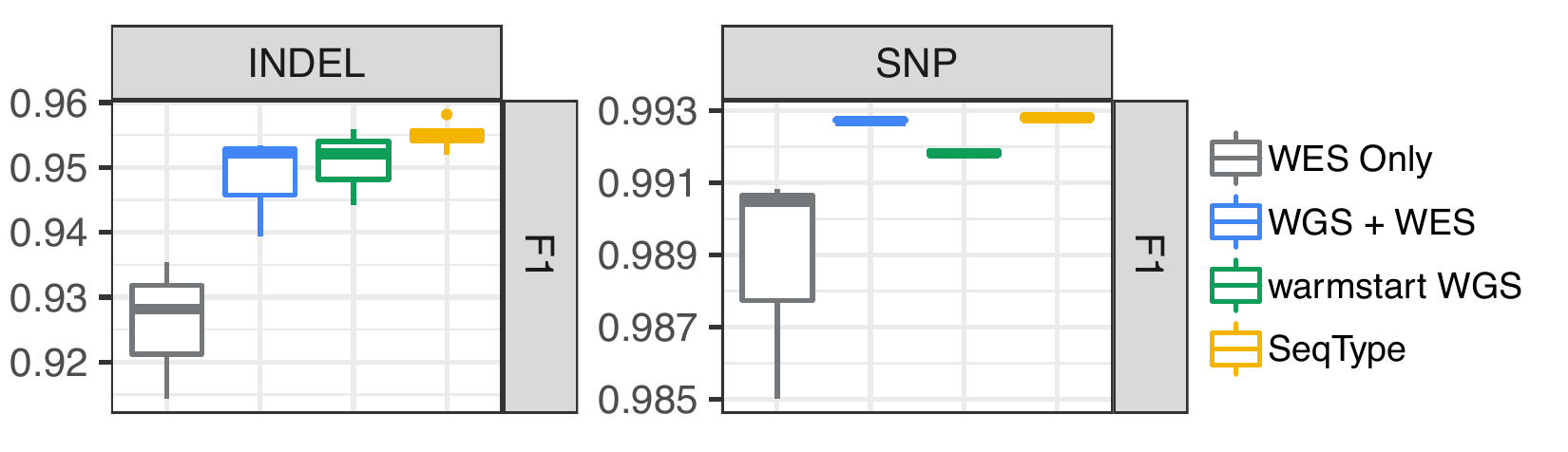}
\caption{Model performances on indels and SNPs, evaluated on the whole exome sample HG002.}
\label{fig:data_aug}
\end{figure}

We first evaluate two strategies--\textit{WGS + WES} and \textit{warmstart WGS}--for adding training examples from WGS relative to a \textit{WES only} baseline (Figure~\ref{fig:data_aug}). Both strategies improve DeepVariant F1 scores ($p_{\text{WGS + WES}} = 3.3 \times 10^{-173}$, $p_{\text{warmstart WGS}} = 5.4 \times 10^{-153}$). Additionally, the addition of WGS data reduces the variability in model performance across replicated experiments. 

\begin{wrapfigure}{r}{0.5\textwidth}
     \includegraphics[width=\linewidth]{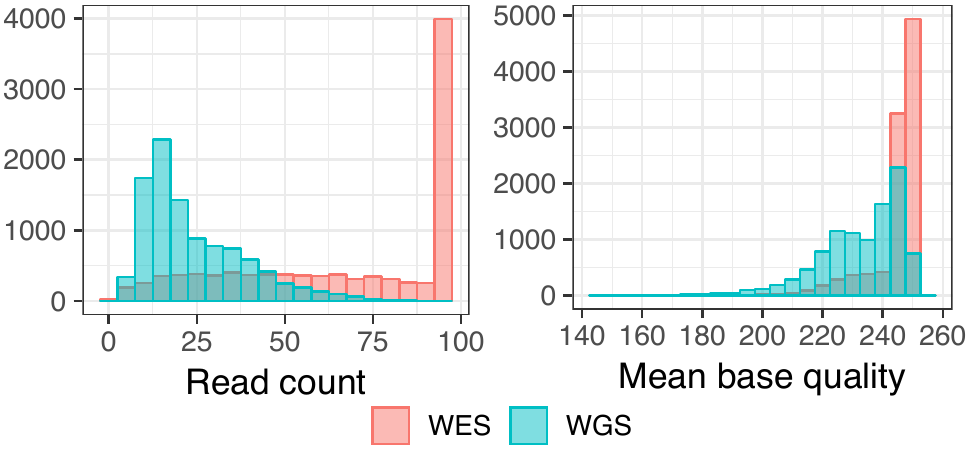}
     \caption{Histograms of read count and mean base quality per example, collected from 10k DeepVariant examples.}
     \label{fig:data_distribution}
\end{wrapfigure}

We stratify performance by variant type. SNPs are substitutions that do not change the sequence length, while indels introduce insertions or deletions. Indel variants are harder to accurately predict (\citep{giab_data}, Figure~\ref{fig:single_datatype}), especially in WES due to additional biases in coverage of GC-rich and poor regions~\citep{wgsBetterWes}. Plotting sequence features of WGS and WES examples reveals the differences between these sequencing types (Figure~\ref{fig:data_distribution}).

We then evaluate DeepVariant performance after adding a sequencing type feature vector with 200 dimensions (Figure~\ref{fig:data_aug}). The \textit{SeqType} model is trained on \textit{WGS + WES} data configuration. \textit{SeqType} significantly improves indels and SNPs F1 scores as opposed to \textit{WES only} ($p_{\text{SeqType}} < 4.1 \times 10^{-288}$). Compared with three other methods, \textit{SeqType} reduces the total number of prediction errors by 
6\% - 38\% on indels (\textit{WGS + WES}: 6\%; \textit{warmstart WGS}: 13\%; and \textit{WES only}: 38\%), and 0.74\% - 36\% on SNPs (\textit{WGS + WES}: 0.74\%; \textit{warmstart WGS}: 12\%; and \textit{WES only}: 36\%). We also note a further reduction in the variability of trained model accuracy on indels.

\begin{figure}[!ht]
\centering
\includegraphics[width=0.95\textwidth]{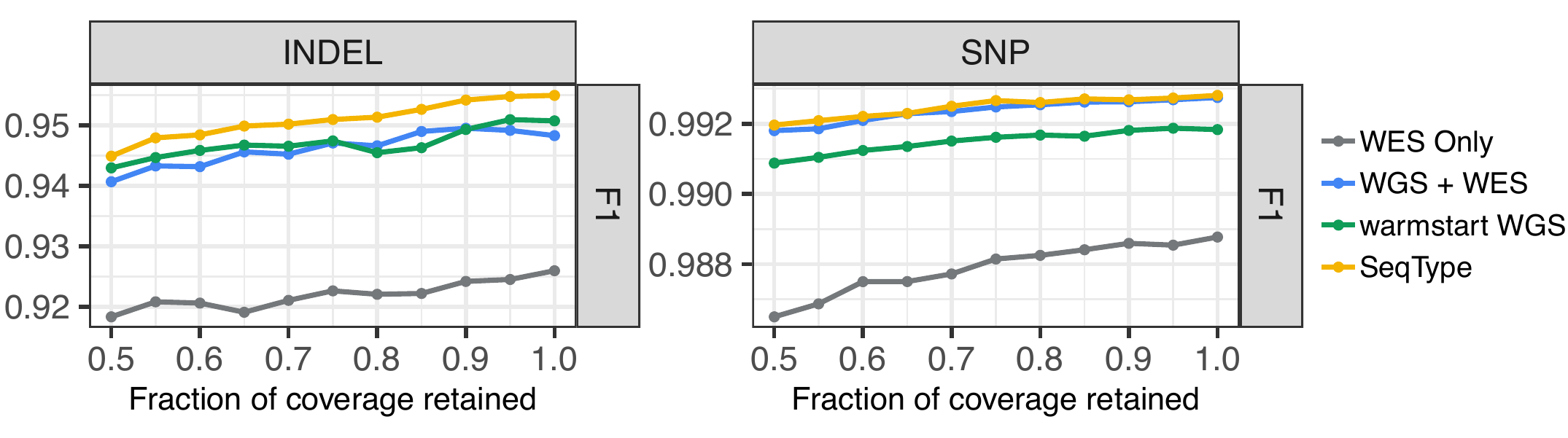}
\caption{Model performances across with different fractions of coverage retained.}
\label{fig:downsample}
\end{figure}

We further measure performance of each model on progressively harder test sets by randomly downsamping the coverage of the WES samples (Figure~\ref{fig:downsample}). \textit{WGS + WES} and \textit{warmstart WGS} both outperform \textit{WES only}. \textit{WGS + WES} shows higher SNPs F1 scores across all downsample fractions tested, whereas both \textit{WGS + WES} and \textit{warmstart WGS} remain roughly the same for indels. Adding the sequencing type feature further improves indels F1 scores, while matching or slightly improving SNPs F1 measures. This result is consistent across all downsample fractions.

%% file: conclusion.tex
Variant calling has become increasingly beneficial for research and clinical diagnoses~\citep{yang2014molecular,xue2015solving}. Here we present three data augmentation strategies to improve genetic variant calling from WES data. We show that incorporating WGS data during training by 1) jointly training on WGS and WES data, and 2) warmstarting the WES model from a WGS model improve accuracy on WES data. Since WGS and WES data come from different distributions, we observe further improvements by 3) jointly training on WGS and WES data and including the sequencing type information through a low-dimensional feature vector. This approach shows the most improvement on indels. All three approaches are robust to downsampling and perform well on lower-coverage data.

The sequencing type information can be encoded using fewer dimensions and does not necessarily need to be learned. We experiment with two other variations of the \textit{SeqType} method: 1) trainable vectors of 100 dimensions, and 2) replacing the trainable vectors with constant vectors, where all values are 0 for WGS data or 1 for WES data. Our preliminary results suggest neither of these attempts successfully improves prediction accuracy. These observations indicate it is beneficial to use trainable vectors to distinguish sequencing types as these vectors can potentially learn to encode unique sequencing type features.

The \textit{SeqType} method naturally extends from the concept of embeddings, which refer to a set of representation techniques commonly used in natural language processing~\citep{word2vec,GTP-2,BERT,roberta} and genomics~\citep{biovec,deepNF,bindSpace}. Unlike other embedding methods which focus on dimension reduction, \textit{SeqType} vector embeddings are trained to learn abstract features of their corresponding data types. We believe this method can be readily applied to other data augmentation problems. For instance, variant callers trained on Illumina NGS data may not generalize well to Pacific Biosciences data due to their vastly different sequencing and error profiles~\citep{opportunities}. Despite both being Illumina high-capacity sequencers, HiSeq and NovaSeq reads have noticeably different alignment characteristics. We hypothesize learning sequencer-specific embeddings will be particularly useful in these scenarios, as the embeddings can potentially capture features unique to each sequencing platform.